\documentclass[aps,prb,twocolumn,superscriptaddress,floatfix,showpacs]{revtex4}

\usepackage{graphicx}
\usepackage{dcolumn}
\usepackage{amsmath}
\usepackage{gensymb}
\usepackage{hyperref}
\usepackage[usenames]{color}

\begin{document}

\title{Electronic correlations and crystal structure distortions in  BaBiO$_3$}

\author{Dm.~Korotin}
\affiliation{Institute of Metal Physics, S.~Kovalevskoy St. 18, 620990 Yekaterinburg, Russia}
\author{V.~Kukolev}
\affiliation{Ural State University, Pr.~Lenina 51, 620083 Yekaterinburg, Russia}
\author{A.~V.~Kozhevnikov}
\affiliation{Institute for Theoretical Physics, ETH Z\"urich, CH-8093 Z\"urich, Switzerland}
\author{D.~Novoselov}
\affiliation{Institute of Metal Physics, S.~Kovalevskoy St. 18, 620990 Yekaterinburg, Russia}
\author{V.~I.~Anisimov}
\affiliation{Institute of Metal Physics, S.~Kovalevskoy St. 18, 620990 Yekaterinburg, Russia}
\affiliation{Ural Federal University, 620002 Yekaterinburg, Russia}

\begin{abstract}
BaBiO$_3$ is a material where formally Bi$^{4+}$ ions with the half-filled $6s$-states form the alternating set of Bi$^{3+}$ and Bi$^{5+}$ ions resulting in a charge ordered insulator. The charge ordering is accompanied by the breathing distortion of the BiO$_6$ octahedra (extension and contraction of the Bi-O bond lengths). Standard Density Functional Theory (DFT) calculations fail to obtain the crystal structure instability caused by the pure  breathing distortions. Combining effects of the breathing distortions and tilting of the BiO$_6$ octahedra allows DFT to reproduce qualitatively experimentally observed insulator with monoclinic crystal structure but gives strongly underestimate breathing distortion parameter and energy gap values. In the present work we reexamine the BaBiO$_3$ problem within the GGA+U method using a Wannier functions basis set for the Bi $6s$-band. Due to high oxidation state of bismuth in this material the Bi $6s$-symmetry Wannier function is predominantly extended spatially on surrounding oxygen ions and hence differs strongly from a pure atomic $6s$-orbital. That is in sharp contrast to transition metal oxides (with exclusion of high oxidation state compounds) where the major part a of $d$-band Wannier function is concentrated on metal ion and a pure atomic $d$-orbital can serve as a good approximation. 
The GGA+U calculation results agree well with experimental data, in particular with experimental crystal structure parameters and energy gap values. Moreover, the GGA+U method allows one to reproduce the crystal structure instability due to the pure  breathing distortions without octahedra tilting. 
\end{abstract}

\maketitle
\section{Introduction}
BaBiO$_3$ is a parent compound for high-$T_c$ superconductors Ba$_{1-x}$K$_{x}$BiO$_3$ and
Ba$_{1-x}$Pb$_{x}$BiO$_3$. 
Properties of these compounds are the most intriguing when  chemical composition  is near transition to SC state.
It would be promising to describe softening of phonon modes with doping as a result of 
{\em ab initio} calculation in order to support phonon mechanism for superconductivity 
in these materials. The proper description of electronic 
and crystal structure of pure BaBiO$_3$ has serious difficulties within Density Functional Theory (DFT).
A possible solution to this problem is to take into account strong electronic correlations that are important in high-$T_c$ cuprates and
oxipnictides. However, the most straightforward way to include correlation effects: the LDA+U method fails in attempts to improve results of the DFT calculation for BaBiO$_3$. In the present work it is shown that the reason for this failure is a high oxidation state of the compound. A conventional basis set used in the LDA+U method is pure atomic orbitals, that is usually a good approximation for $d$-bands Wannier functions in transition metal compounds (with important exception of high oxidation state materials). For BaBiO$_3$ one should use genuine Wannier functions as a basis set for the LDA+U correction potential, because they differ a lot from pure atomic Bi $6s$-orbitals. Below it is demonstrated that the LDA+U correction in Wannier function basis results in a strong improvement of the calculated parameters for crystal structure and spectral properties for BaBiO$_3$.

The LDA+U~\cite{Anisimov1991} and LDA+DMFT~\cite{Held2006} methods  for the electronic structure 
calculations of strongly correlated systems are designed as a combination of DFT and model Hamiltonian approach (such as Hubbard or Anderson models).  A DFT potential is defined by electronic density only and hence is invariant under unitary transformation of auxiliary Kohn-Sham orbitals (DFT knows nothing about basis functions of a particular method; DFT knows only Kohn-Sham orbitals and total charge density). However, in model Hamiltonians Coulomb interaction between electrons is explicitly defined for localized states having symmetry of atomic orbitals.  A specific form of these orbitals remains to be determined in a calculation scheme developed for LDA+U method realization. For the LMTO~\cite{Andersen1975} method, the basis set (Muffin-tin orbitals) is explicitly constructed in the form of atomic-like orbitals and so in this case the choice of orbitals for Coulomb interaction term is obvious. For the LAPW\cite{Singh1994} method, there are atomic symmetry wave functions inside muffin-tin spheres and there are pseudoatomic orbitals even for pseudopotential methods. These orbitals are usually used to define the Coulomb interaction term in a LDA+U calculation scheme based on corresponding methods. 

The most general and ``rigorous'' way to define atomic-like orbitals in solid is to build Wannier functions. That is done via Fourier transformation from delocalized Bloch functions $\psi_{\bf k}$ in reciprocal space of wave vectors $\bf k$ to  functions $W^{\bf T}$ in direct space lattice sites vectors $\bf T$. Both sets,  the Bloch functions $\psi_{\bf k}$ and Wannier  functions $W^{\bf T}$, generate the same wave functions Hilbert space and can be considered as different basis sets for this space. Correlation effects are usually strong enough to be taken into account for transition metal compounds, where $d$- or $f$-states produce narrow energy bands. Fourier transformation for corresponding Bloch functions results in  Wannier functions, which are well localized inside an atomic sphere and hence using atomic $d$- or $f$-orbitals instead of them is a good approximation.

Correlated states differ substantially from  pure atomic $d$-orbitals in compounds
with elements in high oxidation state, for example, CaFeO$_3$, AuTe$_2$, and BaBiO$_3$. 
Narrow energy bands near the Fermi level have a serious admixture of $p$-states in such compounds, which could be even larger in value than $d$-states contribution.
Moreover, an energy gap could be formed predominantly by $p$-states with a slight admixture of  $d$-states 
(as it happens in AuTe$_2$)~\cite{BCHKrutzenandJEInglesfield1990}. The mixed nature of correlated 
electronic states does not allow to use traditional implementations of the LDA+U  method with pure atomic orbitals as a basis set for Coulomb interaction term. Explicit Wannier functions basis set for 
correlated states should be chosen to perform electronic structure calculations for these materials.

Many researchers used Wannier function basis for strongly correlated materials calculations in recent years, for example, see Refs.~\cite{Leonov2008,Pavarini2004,Yin2006,Lechermann2006}. In the present work we investigate BaBiO$_3$ where Wannier functions basis in contrast to atomic orbitals is crucial for proper crystal structure description. In the case of BaBiO$_3$ partially filled  band, which is a source of correlation effects, is formed by Bi $6s$-states and oxygen $2p$-states. The energy of original Bi $6s$-level is below oxygen band but strong hybridization between them results in the splitting of the partially filled band upward from the $2p$-states manifold. The contribution of oxygen states to this band is stronger then contents of Bi $6s$-states but corresponding wave function still has $s$-symmetry in respect to Bi site. Wannier function constructed from this band has the same symmetry but it is spatially distributed more on oxygen sites than on Bi ion. Accounting for the Coulomb interaction on such a Wannier functions basis leads to the results very different from the LDA+U calculations using pure Bi $6s$-orbitals.

\section{Calculation method}
\subsection{Wannier functions}
Wannier functions (WFs) $| W_n^{\bf T} \rangle$ are defined as Fourier transforms 
of the Bloch functions $| \Psi_{n{\bf k}} \rangle$: 
\begin{equation}
\label{Wannier:Definition}
| W_n^{\bf T} \rangle = \frac{1}{\sqrt{\Omega}} \sum_{\bf k} e^{-i{\bf kT}}| \Psi_{n{\bf k}} \rangle,
\end{equation}
where ${\bf T}$ is the lattice translation vector, $n$ is the band number, and ${\bf k}$ is
the wave vector.
WFs seem to be an optimal choice to describe correlated states because these functions are
localized, site-centered and represent a complete basis set for the 
Bloch functions Hilbert space.
In the present paper WFs were generated as projections of Bloch sums of the
atomic orbitals $|\phi_{n{\bf k}}\rangle = \sum_{\bf T} e^{i{\bf kT}}|\phi_n^{\bf T}\rangle $ onto a subspace of the Bloch functions (the detailed description 
 of WFs construction procedure within pseudopotential method is provided in Ref.~\cite{Korotin2008a}):
\begin{equation}
|W^{\bf T}_n\rangle = \frac{1}{\sqrt{N_{\bf{k}}}} \sum_{\bf k} |W_{n{\bf k}}\rangle e^{-i{\bf kT}},
\end{equation}
\begin{equation}
\label{Wannier:proj}
|W_{n{\bf k}}\rangle \equiv \sum_{i=N_1}^{N_2} |\Psi_{i{\bf k}}\rangle\langle \Psi_{i{\bf k}} | \phi_{n{\bf k}} \rangle.
\end{equation}

The resulting WFs have symmetry of the atomic orbitals $\phi_n$ and describe electronic states that
form energy bands numbered from $N_1$ to $N_2$. To describe Coulomb repulsion between 
occupied and empty states one needs to compute occupation numbers of WFs.
The Wannier functions occupancy matrix $Q^{WF}_{nm}$ is given by
\begin{equation}
\label{Wannier:Q}
Q^{WF}_{nm} = \langle W^0_n | \left ( \sum \limits_{\bf k} \sum \limits_{i=N_1}^{N_2} 
|\Psi_{i{\bf k}} \rangle \theta(\varepsilon_i({\bf k}) - E_f ) 
\langle \Psi_{i{\bf k}} | \right ) | W^0_m \rangle 
\end{equation}
where $\theta$ is the step function, $\varepsilon_i({\bf k})$ is the one-electron energy for the 
state $i$, and $E_f$ is the Fermi energy.


\subsection{LDA+U method in Wannier functions basis}
One of the most general and accurate approaches to describe the electronic structure of strongly 
correlated systems is the
 dynamical mean-field theory (DMFT)~\cite{D.Vollhardt,Pruschke1995,Kotliar2006}.
The method combining {\em ab initio} DFT approach and model DMFT calculations -- LDA+DMFT
was applied recently to the quantitative description of both electronic and crystal structure of several
compounds with strong electronic correlations~\cite{Leonov2009,Wilson1971,Kunes2009,Leonov2008,Leonov2011}.
The LDA+DMFT method is successful and promising but it demands a lot of computational resources. 
In the present paper the static limit of Dynamical Mean-Field Theory --  the LDA+U method 
is used. There is an additional term in Hamiltonian operator to take into account Coulomb correlations. If one considers on-site interaction only and takes into account 
the diagonal form of the occupation matrix $Q^{WF}_{nm}$, the additional term is written in WFs basis as
\begin{equation}
\label{WannierU}
\hat H^\prime ({\bf k}) =  \sum_{m} |W_{m{\bf k}} \rangle 
\left( U \times (\frac{1}{2}-n_{m}) \right) \langle W_{m{\bf k}} |,
\end{equation}
where $n_{m}=Q^{WF}_{mm}$ -- the occupation number for {\em m}-th WF. In the case of BaBiO$_3$ the non-diagonal terms of $Q^{WF}_{nm}$ are equal to zero by construction (we have strictly orthogonal WFs of Bi-s-orbital symmetry). For more complex basis sets it could be necessary to use rotationally invariant potential correction and take into account non-diagonal terms.

The full Hamiltonian operator is written as
\begin{equation}
\hat H^{LDA+U} = \hat H^{LDA} + \hat H^\prime,
\end{equation}
where $\hat H^{LDA}$ -- the Hamiltonian operator within LDA (or GGA) approximation. The form of the correction potential (\ref{WannierU}) results in a negative addition to potential $-U/2$ for occupied states $n_{m}=1$ and in a positive value $+U/2$ for empty states $n_{m}=0$. Hence the LDA+U correction to DFT will increase energy separation between bands below and above the Fermi level increasing energy gap values and enhancing total energy gain of the distortion that causes the gap appearance. 

The value of Coulomb interaction parameter $U$ is computed via constrained LDA calculation as described 
in~\cite{Korotin2008a}.

The total energy is computed as
\begin{equation}
	E^{tot} = E^{LDA} + E^{U} - E^{DC},
\end{equation}
where $E^{LDA}$ is the total energy from a standard DFT calculation (LDA or GGA could be used, we preferred GGA), $E^{U} = \frac{1}{2}\sum_{m\neq m\prime}Un_mn_m\prime$, $E^{DC} = \frac{1}{2}Un(n-1)$ is the double counting correction, and $n=Tr(Q^{WF}_{nm})$ is the total occupancy of WFs.


\section{Results and discussion}
 
Undoped BaBiO$_3$ has a monoclinic crystal structure (symmetry group is C2/m). The structure could be obtained from an ideal cubic perovskite structure
simultaneous breathing distortion of the BiO$_6$
octahedra and tilting the octahedra around the [110] axis. 
Primitive cell contains two formula units. Chemical formula could be written~\cite{Cox1976969} 
as Ba$_2^{2+}$Bi$^{3+}$Bi$^{5+}$O$_6^{2-}$. Two different lengths of Bi-O bonds correspond to 
two different valences of Bi ion: the short to Bi$^{5+}$-O and the long to  Bi$^{3+}$-O. 
The neighboring Bi-type ions for Bi$^{3+}$ are Bi$^{5+}$ and vice versa. The alternation of Bi$^{3+}$
and Bi$^{5+}$ ions forms charge density wave. Experimental value~\cite{Cox1976969} for the breathing distortion is
$b$=0.085~$\AA$  and for the tilting is $t$=10.3\degree. Spectroscopy measurements show that BaBiO$_3$ is an insulator 
with the energy gap value
 $\approx$0.5~eV\cite{TakagiHUchidaSTajimaS1987}.

Liechtenstein {\em et al.} performed calculations of electronic and crystal structure of BaBiO$_3$
with full-potential LMTO method~\cite{PhysRevB.44.5388}. The calculated values of the BiO$_6$ octahedra
distortion ($b$=0.055~\AA, $t$=8.5\degree) underestimate the experimental data as well as the energy gap value $\approx$0.02~eV. Also authors were unable to 
obtain cubic crystal structure instability against pure breathing octahedra distortion without tilting.

Franchini {\em et al.} have modeled the crystal structure distortions more 
successfully~\cite{Franchini2009,Franchini2010}. The authors used Heyd-Scuseria-Ernzerhof (HSE) 
hybrid functional that includes  25\% of exact Hartree-Fock exchange in addition to DFT, so it partially takes into account correlation effects. The hybrid 
functional allowed to reproduce energy gap value with slight overestimation (0.65 eV) and
get octahedra distortion values close to experimental ones ($b$=0.09~\AA, $t$=11.9\degree).

\begin{figure}[ht!]
\centerline{\includegraphics[width=0.45\textwidth]{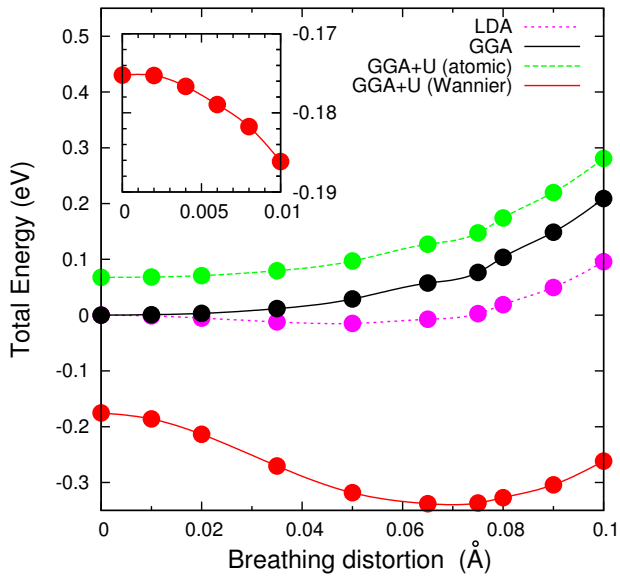}}

\caption{(color online) Calculated within LDA (magenta dashed), GGA (black solid), atomic orbitals GGA+U (green dashed), and GGA+U in WFs basis (red solid) total energy of BaBiO$_3$ as a function of pure breathing for tilting = 0\degree. The inset shows results of GGA+U in WFs basis calculations close to zero breathing distortion.}
\label{fig:emin0}
\end{figure}

Thonhauser and Rabe ~\cite{Thonhauser2006} have reported the cubic crystal structure instability in respect to breathing distortion of the BiO$_6$ octahedra using obsolete Local Density Approximation ( LDA) while more rigorous Generalized Gradient Approximation (GGA) calculations do not show any minimum at all (see Fig. \ref{fig:emin0}).
The total energy lowering value for distorted structure in LDA calculations  ~\cite{Thonhauser2006} was 3.35~meV per formula unit 
corresponding to crystal structure transformation temperature as low as 30~K, that clearly contradicts to experimental fact that distorted cubic  crystal structure of BaBiO$_3$ is stable
at room temperature and higher temperatures~\cite{PhysRevB.41.4126}.

In the present work to describe the correlated states with WFs, a self-consistent calculation
within the generalized gradient approximation (GGA) was performed as the first step. The pseudopotential plane-waves method implemented
in the Quantum-ESPRESSO package~\cite{Giannozzi2009}, was used.  Vanderbilt ultrasoft 
pseudopotentials~\cite{Vanderbilt1990} were taken from QE pseudopotentials library.
A kinetic-energy cutoff for the plane-wave expansion of the electronic states was set to 45 Ry. 
Integrations in reciprocal space were performed using (10,10,10) Monkhorst-Pack~\cite{Monkhorst1976} 
k-point grid in the full Brillouin  zone.

\begin{figure}[tbp!]
\centerline{\includegraphics[width=0.49\textwidth]{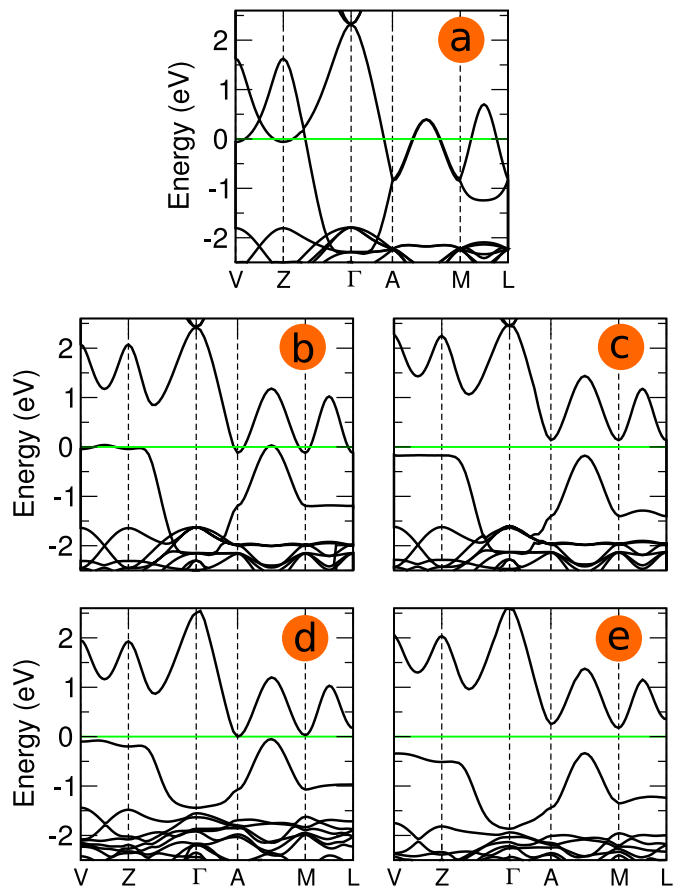}}

\caption{Band structure of BaBiO$_3$ obtained for various crystal structures within the GGA and GGA+U approaches. ({a\label{fig:5bands:a}}) Ideal cubic crystal structure, GGA. ({b\label{fig:5bands:b}}) Distorted crystal structure with the breathing-only distortions, $b$=0.075~$\AA$, GGA. ({c\label{fig:5bands:c}}) Distorted crystal structure with the breathing-only distortions, $b$=0.075~$\AA$, GGA+U in WFs basis. ({d\label{fig:5bands:d}}) Monoclinic crystal structure ($b$=0.075 \AA, $t$=12\degree), GGA. ({e\label{fig:5bands:e}}) Monoclinic crystal structure ($b$=0.075 \AA, $t$=12\degree), GGA+U in WFs basis. Zero energy corresponds to the Fermi level.}
\label{fig:5bands}
\end{figure}

The GGA calculation results in semiconducting solution with a very small indirect energy gap $\approx$0.05~eV. Two energy bands in the interval [-1.5; 2.5]~eV, see Fig.\ref{fig:5bands},  are generally 
formed by the O-$p$ states, but they have symmetry of the Bi-$6s$ states. The Bi-$5d$ and
Bi-$6p$ partial densities of states give vanishing contribution to the states near the Fermi level.

\begin{figure}[tbp!]
\centerline{\includegraphics[width=0.3\textwidth,clip]{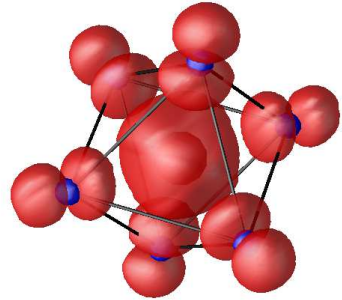}}

\caption{(color online) Squared moduli of the Wannier function centered on Bi$^{3+}$ ion (a big red sphere in the center). The blue spheres are oxygen ions.}
\label{fig:wannier}
\end{figure}

WFs were generated by projection of two Bi-$s$ atomic orbitals onto subspace defined by two energy bands near the Fermi level. Squared moduli of WF centered on Bi$^{3+}$ ion is shown in Fig.~\ref{fig:wannier}. WF has symmetry of $s$-orbital. However, it has significant contribution from $p$-states of neighboring oxygen ions. The second
WF (centered on Bi$^{5+}$ ion) will differ only by the ratio between O-$p$
and Bi-$s$ states. 

The contribution of oxygen states to partially filled bands could be accounted for by including in the Hamiltonian  Hubbard $U$ correction to the oxygen $p$-states in addition to Bi-$s$ orbitals. However 
 in that case one needs to take into account also off-diagonal inter-atomic $s-p$ terms in occupation matrix. The WF approach proposed in the present work is simpler.

To perform calculation within LDA+U method in WFs basis, the $U$ value for BaBiO$_3$ was computed via the constrained LDA method. The Hubbard $U$ value could be calculated as the second derivative of total energy over occupation of correlated state~\cite{Anisimov1991b}:
\begin{equation} \label{UD2}
U = \frac{\partial^2 E_{DFT}}{\partial^2 n_{corr}}.
\end{equation}
In DFT one-electron energy of a state is a derivative of total energy in respect to the state occupancy. Then $U$ value can be calculated as:
\begin{equation} \label{U1}
U = \frac{\partial \epsilon_{corr}}{\partial n_{corr}},
\end{equation}
where $\epsilon_{corr}$ is one particle energy of correlated state, and $n_{corr}$ -- its occupation number. The partial derivative is taken numerically as described below.
Constrained potential that affects only states considered as correlated ones is defined as a projection operator:
\begin{equation} \label{HConstr2}
\hat H^{\bf{k}}_{constr} = \sum_n |W_{n\bf{k}}\rangle \delta V_n\langle W_{n\bf{k}}|,
\end{equation}
where $\delta V_n$ is a small number, for example, $\pm$~0.1~eV, positive for WF centered on one Bi-site and negative for WF centered on another Bi-site. The constrained potential is included into the Hamiltonian operator during self-consistency cycle, that allows to take into account
screening of charge redistribution. As a result, the WFs occupation numbers and energies change from unconstrained results. The derivative~(\ref{U1}) is calculated as a ratio of these variations. We have checked that the calculated derivative value of $U$ does not depend on $\delta V_n$, by using its various values equal to ($0.1, 0.2, 0.3, 0.4~eV$ ).

The resulting $U$ value for BaBiO$_3$ equals
to 0.7~eV. 
The value of Coulomb parameter strongly depends on spacial form of basis function~\cite{Anisimov2008}. Therefore the rather small  value 0.7~eV obtained in our calculations is reasonable for WFs shown on Fig.~\ref{fig:wannier} that is rather extended in space. The atomic Bi-s orbital is much more localized in space than WFs hence the $U$ value for atomic orbitals was found to be significantly larger $U$= 2.5~eV.

The $U$ value for WFs basis is twice smaller than the width of the partially filled energy bands that indicates relative weakness of Coulomb interaction
in BaBiO$_3$. But, as it is shown below, this Coulomb interaction
plays the important role in the formation of the electronic states  and crystal structure of the compound.

A simulation of the oxygen octahedra tilting and breathing distortion was performed in a monoclinic (I2/m) supercell with 2 formula units. Sixty 
primitive cells of Ba$_2^{2+}$Bi$^{3+}$Bi$^{5+}$O$_6^{2-}$ were generated corresponding to different combinations of the breathing
and tilting distortions. For every structure a self-consistent calculation within GGA was done. Then WFs were constructed and self-consistent calculations within LDA+U (actually GGA+U) were performed. Coulomb interaction correction was applied only to the states described with the two WFs as defined in Eq.~(\ref{WannierU}).

In Fig.~\ref{fig:5bands} the formation of the energy gap due to the cubic structure distortion is shown. Band structure of ideal cubic 
BaBiO$_3$, see Fig.~\ref{fig:5bands:a}~(a), is definitely metallic. Two energy bands in the [-1.5; 2.5]~eV energy interval cross 
the Fermi level. These bands are double degenerate along the A$\to$M direction. The cubic crystal structure is unstable against 
distortion that lowers the symmetry and opens an energy gap in band structure.

In BaBiO$_3$ the breathing distortion of the oxygen
octahedra plays the crucial role. In Figs.~\ref{fig:5bands:b}~(b) and (c) band structure is shown for the crystal with the breathing-only
distortions of the oxygen octahedra. The breathing value is $b$=0.075~$\AA$. The degeneracy breaking along the A$\to$M direction is clearly seen
in the GGA calculation result. Two separate bands appear in the [-1.5; 2.5]~eV interval  but the Fermi level crosses both of them and the energy gap does not open. Accordingly there is no total energy minimum  due to the
breathing distortion in the GGA calculation. That is clearly illustrated by the black curve in Fig.~\ref{fig:emin0}. The dependence of total energy of BaBiO$_3$ cell on the breathing distortion calculated within GGA has a minimum for the distortion parameter value $b$=0~$\AA$ (stable cubic structure).  Hubbard interaction correction (GGA+U) in the basis of WFs leads to appearance of the band gap equal to 0.3~eV, Fig.~\ref{fig:5bands:c}~(c). Also cubic crystal
structure of BaBiO$_3$ is unstable against the pure breathing distortion in the GGA+U calculation. The total energy of the cell has a minimum for 
$b$=0.075~$\AA$, see Fig.~\ref{fig:emin0}. The energy lowering is equal to 326 meV per formula unit. 

The total energy dependence calculated with the GGA+U method in atomic orbitals basis is also shown in Fig.~\ref{fig:emin0} with the dashed green curve. The minimum of the total energy corresponds to the cubic crystal structure. One can see that atomic orbitals basis potential correction calculation does not result in an improvement of crystal structure description. 
The energy dependence obtained with the LDA calculation is shown on Fig.~\ref{fig:emin0} with the dashed magenta curve. 
It has very shallow minimum for non zero breathing distortion in agreement with~\cite{Thonhauser2006}, but there is no any minimum in the curve obtained in
 GGA that is supposed to be more accurate approximation than LDA.

\begin{figure}[tbp!]
\centerline{\includegraphics[width=0.45\textwidth]{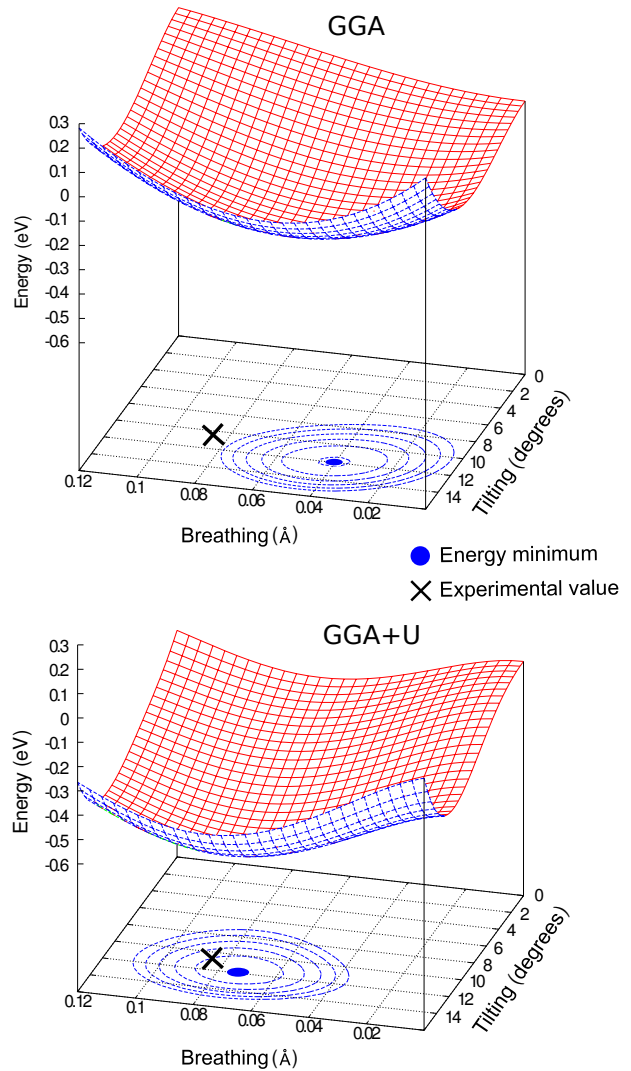}}

\caption{(color online) Total energy of BaBiO$_3$ as a function of breathing and tilting distortions calculated within GGA and GGA+U. Zero energy level corresponds to cubic structure ($b$=0, $t$=0) within GGA. Experimental minimum \cite{PhysRevB.41.4126,Cox1976969} is located at ($b$=0.085 \AA, $t$=10.3\degree) and is marked by the cross; The GGA energy minimum value is at ($b$=0.04 \AA, $t$=12\degree) and the GGA+U value is at ($b$=0.075 \AA, $t$=12\degree)}
\label{fig:emin}
\end{figure}

In the monoclinic cell of BaBiO$_3$ with simultaneous breathing and tilting of BiO$_6$ octahedra the energy gap (equal
to 0.05~eV)  exists in the result of the GGA
calculation, see Fig.~\ref{fig:5bands:d}~(d). Despite the fact that the gap value is very small the cubic structure is unstable against combined distortion in GGA. For the monoclinic structure the total energy minimum corresponds to the distortion parameters values
($b$=0.04 \AA, $t$=12\degree). The values underestimate experimental data in agreement with the previous 
calculations~\cite{PhysRevB.44.5388}.

The GGA+U correction for monoclinic crystal cell increases the energy gap value to 0.55~eV, see Fig.~\ref{fig:5bands:e}~(e), in a good agreement with the experimental value of 0.5~eV.

The splitting of the half-filled double degenerate Bi-$s$ band and the opening of the energy gap results in a filled band corresponding to the $6s$-state of one Bi-site (Bi$^{3+}$) 
and an empty band formed by the states of another Bi-site (Bi$^{5+}$). A charge difference between these two Bi-sites (equal to 2$e$) is expected for fully ionic picture. 
However, the occupancy values difference for  two atomic Bi-$6s$ orbitals  in the LDA+U calculation is equal to 0.43$e$ only in agreement with \cite{Franchini2009}. 
The small value of the charge difference can be understood from an analysis of WFs contents. The Bi-6$s$ symmetry Wannier function for Bi$^{3+}$ site has only 23\% of Bi-$s$ atomic orbital 
and 77\% of O-$p$ orbitals; on Bi$^{5+}$ site WF composition is 20\% of Bi-$s$ and 80\% of O-$p$. A change in occupation number of WF by $\pm$1 results 
in a change of $s$ atomic orbital occupancy by +0.23 for Bi$^{3+}$ and -0.20 for Bi$^{5+}$, that gives 0.43 in total.

In Fig.~\ref{fig:emin} the total energy of  BaBiO$_3$ in GGA and GGA+U as a function of breathing and tilting distortions is shown. Zero energy value corresponds to the GGA total energy of ideal cubic
crystal structure. The surface was obtained by extrapolation from 60 points as the 4$^{th}$ order polynomial. The experimental values of the distortions are ($b$=0.085~\AA, $t$=10.3\degree) and is marked by the cross in Fig.~\ref{fig:emin}. The GGA calculations gave total energy minimum  for distortion value parameters ($b$=0.04~\AA, $t$= 12\degree)  strongly underestimating the experimental data. The total energy minimum within the
GGA+U calculation corresponds to distortion parameter values ($b$=0.075~\AA, $t$=12\degree) that are very close to the experimental ones.

Figure~\ref{fig:emin} clearly illustrates the effectiveness of the calculation scheme proposed in the present paper. The usage of the WFs basis
for the partially filled states in the GGA+U method allows one to reproduce not only the energy gap value, but also noticeably improve the description of the crystal structure distortions in BaBiO$_3$. The most important result is the instability of cubic crystal against the pure breathing distortion obtained in the GGA+U calculations. 

\section{Conclusion}
The scheme for studying of electronic correlations in high oxidation stated compounds based on the LDA+U method in Wannier functions basis was proposed and applied to the case of BaBiO$_3$. The energy gap value and crystal structure parameters of monoclinic BaBiO$_3$ were successfully 
described in agreement with experimental data. Instability of cubic crystal structure in respect to pure the breathing distortion was obtained for the first time. 

\begin{acknowledgments}
Support by the Russian Foundation for Basic Research under grants no. RFFI-10-02-00046a, RFFI-10-02-96011ural, President of Russian Federation Fund of Support for Scientific Schools grant 1941.2008.2 and the Program of the
Russian Academy of Science Presidium ''Quantum physics
of condensed matter'' is gratefully acknowledged. Dmitry Korotin acknowledges the support of Dynasty Foundation.
\end{acknowledgments}


\begin{thebibliography}{29}
\expandafter\ifx\csname natexlab\endcsname\relax\def\natexlab#1{#1}\fi
\expandafter\ifx\csname bibnamefont\endcsname\relax
  \def\bibnamefont#1{#1}\fi
\expandafter\ifx\csname bibfnamefont\endcsname\relax
  \def\bibfnamefont#1{#1}\fi
\expandafter\ifx\csname citenamefont\endcsname\relax
  \def\citenamefont#1{#1}\fi
\expandafter\ifx\csname url\endcsname\relax
  \def\url#1{\texttt{#1}}\fi
\expandafter\ifx\csname urlprefix\endcsname\relax\def\urlprefix{URL }\fi
\providecommand{\bibinfo}[2]{#2}
\providecommand{\eprint}[2][]{\url{#2}}

\bibitem[{\citenamefont{Anisimov et~al.}(1991)\citenamefont{Anisimov, Zaanen,
  and Andersen}}]{Anisimov1991}
\bibinfo{author}{\bibfnamefont{V.~I.} \bibnamefont{Anisimov}},
  \bibinfo{author}{\bibfnamefont{J.}~\bibnamefont{Zaanen}}, \bibnamefont{and}
  \bibinfo{author}{\bibfnamefont{O.~K.} \bibnamefont{Andersen}},
  \bibinfo{journal}{Physical Review B} \textbf{\bibinfo{volume}{44}},
  \bibinfo{pages}{943} (\bibinfo{year}{1991}), ISSN \bibinfo{issn}{0163-1829},
  \urlprefix\url{http://link.aps.org/doi/10.1103/PhysRevB.44.943}.

\bibitem[{\citenamefont{Held et~al.}(2006)\citenamefont{Held, Nekrasov, Keller,
  Eyert, Bl\"{u}mer, McMahan, Scalettar, Pruschke, Anisimov, and
  Vollhardt}}]{Held2006}
\bibinfo{author}{\bibfnamefont{K.}~\bibnamefont{Held}},
  \bibinfo{author}{\bibfnamefont{I.~A.} \bibnamefont{Nekrasov}},
  \bibinfo{author}{\bibfnamefont{G.}~\bibnamefont{Keller}},
  \bibinfo{author}{\bibfnamefont{V.}~\bibnamefont{Eyert}},
  \bibinfo{author}{\bibfnamefont{N.}~\bibnamefont{Bl\"{u}mer}},
  \bibinfo{author}{\bibfnamefont{A.~K.} \bibnamefont{McMahan}},
  \bibinfo{author}{\bibfnamefont{R.~T.} \bibnamefont{Scalettar}},
  \bibinfo{author}{\bibfnamefont{T.}~\bibnamefont{Pruschke}},
  \bibinfo{author}{\bibfnamefont{V.~I.} \bibnamefont{Anisimov}},
  \bibnamefont{and}
  \bibinfo{author}{\bibfnamefont{D.}~\bibnamefont{Vollhardt}},
  \bibinfo{journal}{Physica Status Solidi (B)} \textbf{\bibinfo{volume}{243}},
  \bibinfo{pages}{2599} (\bibinfo{year}{2006}), ISSN \bibinfo{issn}{0370-1972},
  \urlprefix\url{http://doi.wiley.com/10.1002/pssb.200642053}.

\bibitem[{\citenamefont{Andersen}(1975)}]{Andersen1975}
\bibinfo{author}{\bibfnamefont{O.~K.} \bibnamefont{Andersen}},
  \bibinfo{journal}{Physical Review B} \textbf{\bibinfo{volume}{12}},
  \bibinfo{pages}{3060} (\bibinfo{year}{1975}), ISSN \bibinfo{issn}{0556-2805},
  \urlprefix\url{http://link.aps.org/doi/10.1103/PhysRevB.12.3060}.

\bibitem[{\citenamefont{Singh}(1994)}]{Singh1994}
\bibinfo{author}{\bibfnamefont{D.}~\bibnamefont{Singh}},
  \emph{\bibinfo{title}{{Planewaves, pseudopotentials and the LAPW method}}}
  (\bibinfo{publisher}{Kluwer Academic Publishing}, \bibinfo{year}{1994}), ISBN
  \bibinfo{isbn}{0-7923-9421-7}.

\bibitem[{\citenamefont{Krutzen and
  Inglesfield}(1990)}]{BCHKrutzenandJEInglesfield1990}
\bibinfo{author}{\bibfnamefont{B.~C.~H.} \bibnamefont{Krutzen}}
  \bibnamefont{and} \bibinfo{author}{\bibfnamefont{J.~E.}
  \bibnamefont{Inglesfield}}, \bibinfo{journal}{Journal of Physics: Condensed
  Matter} \textbf{\bibinfo{volume}{2}}, \bibinfo{pages}{4829}
  (\bibinfo{year}{1990}), ISSN \bibinfo{issn}{0953-8984},
  \urlprefix\url{http://stacks.iop.org/0953-8984/2/i=22/a=005}.

\bibitem[{\citenamefont{Leonov et~al.}(2008)\citenamefont{Leonov, Binggeli,
  Korotin, Anisimov, Stoji\'{c}, and Vollhardt}}]{Leonov2008}
\bibinfo{author}{\bibfnamefont{I.}~\bibnamefont{Leonov}},
  \bibinfo{author}{\bibfnamefont{N.}~\bibnamefont{Binggeli}},
  \bibinfo{author}{\bibfnamefont{D.}~\bibnamefont{Korotin}},
  \bibinfo{author}{\bibfnamefont{V.~I.} \bibnamefont{Anisimov}},
  \bibinfo{author}{\bibfnamefont{N.}~\bibnamefont{Stoji\'{c}}},
  \bibnamefont{and}
  \bibinfo{author}{\bibfnamefont{D.}~\bibnamefont{Vollhardt}},
  \bibinfo{journal}{Physical Review Letters} \textbf{\bibinfo{volume}{101}},
  \bibinfo{pages}{96405} (\bibinfo{year}{2008}), ISSN
  \bibinfo{issn}{0031-9007},
  \urlprefix\url{http://link.aps.org/doi/10.1103/PhysRevLett.101.096405}.

\bibitem[{\citenamefont{Pavarini et~al.}(2004)\citenamefont{Pavarini, Biermann,
  Poteryaev, Lichtenstein, Georges, and Andersen}}]{Pavarini2004}
\bibinfo{author}{\bibfnamefont{E.}~\bibnamefont{Pavarini}},
  \bibinfo{author}{\bibfnamefont{S.}~\bibnamefont{Biermann}},
  \bibinfo{author}{\bibfnamefont{A.}~\bibnamefont{Poteryaev}},
  \bibinfo{author}{\bibfnamefont{A.~I.} \bibnamefont{Lichtenstein}},
  \bibinfo{author}{\bibfnamefont{A.}~\bibnamefont{Georges}}, \bibnamefont{and}
  \bibinfo{author}{\bibfnamefont{O.~K.} \bibnamefont{Andersen}},
  \bibinfo{journal}{Physical Review Letters} \textbf{\bibinfo{volume}{92}},
  \bibinfo{pages}{176403} (\bibinfo{year}{2004}), ISSN
  \bibinfo{issn}{0031-9007},
  \urlprefix\url{http://link.aps.org/doi/10.1103/PhysRevLett.92.176403}.

\bibitem[{\citenamefont{Yin et~al.}(2006)\citenamefont{Yin, Volja, and
  Ku}}]{Yin2006}
\bibinfo{author}{\bibfnamefont{W.-G.} \bibnamefont{Yin}},
  \bibinfo{author}{\bibfnamefont{D.}~\bibnamefont{Volja}}, \bibnamefont{and}
  \bibinfo{author}{\bibfnamefont{W.}~\bibnamefont{Ku}},
  \bibinfo{journal}{Physical Review Letters} \textbf{\bibinfo{volume}{96}},
  \bibinfo{pages}{116405} (\bibinfo{year}{2006}), ISSN
  \bibinfo{issn}{0031-9007},
  \urlprefix\url{http://link.aps.org/doi/10.1103/PhysRevLett.96.116405}.

\bibitem[{\citenamefont{Lechermann et~al.}(2006)\citenamefont{Lechermann,
  Georges, Poteryaev, Biermann, Posternak, Yamasaki, and
  Andersen}}]{Lechermann2006}
\bibinfo{author}{\bibfnamefont{F.}~\bibnamefont{Lechermann}},
  \bibinfo{author}{\bibfnamefont{A.}~\bibnamefont{Georges}},
  \bibinfo{author}{\bibfnamefont{A.}~\bibnamefont{Poteryaev}},
  \bibinfo{author}{\bibfnamefont{S.}~\bibnamefont{Biermann}},
  \bibinfo{author}{\bibfnamefont{M.}~\bibnamefont{Posternak}},
  \bibinfo{author}{\bibfnamefont{A.}~\bibnamefont{Yamasaki}}, \bibnamefont{and}
  \bibinfo{author}{\bibfnamefont{O.~K.} \bibnamefont{Andersen}},
  \bibinfo{journal}{Physical Review B} \textbf{\bibinfo{volume}{74}},
  \bibinfo{pages}{125120} (\bibinfo{year}{2006}), ISSN
  \bibinfo{issn}{1098-0121},
  \urlprefix\url{http://link.aps.org/doi/10.1103/PhysRevB.74.125120}.

\bibitem[{\citenamefont{Korotin et~al.}(2008)\citenamefont{Korotin,
  Kozhevnikov, Skornyakov, Leonov, Binggeli, Anisimov, and
  Trimarchi}}]{Korotin2008a}
\bibinfo{author}{\bibfnamefont{D.}~\bibnamefont{Korotin}},
  \bibinfo{author}{\bibfnamefont{A.~V.} \bibnamefont{Kozhevnikov}},
  \bibinfo{author}{\bibfnamefont{S.~L.} \bibnamefont{Skornyakov}},
  \bibinfo{author}{\bibfnamefont{I.}~\bibnamefont{Leonov}},
  \bibinfo{author}{\bibfnamefont{N.}~\bibnamefont{Binggeli}},
  \bibinfo{author}{\bibfnamefont{V.~I.} \bibnamefont{Anisimov}},
  \bibnamefont{and}
  \bibinfo{author}{\bibfnamefont{G.}~\bibnamefont{Trimarchi}},
  \bibinfo{journal}{The European Physical Journal B}
  \textbf{\bibinfo{volume}{65}}, \bibinfo{pages}{91} (\bibinfo{year}{2008}),
  ISSN \bibinfo{issn}{1434-6028},
  \urlprefix\url{http://www.springerlink.com/index/10.1140/epjb/e2008-00326-3}.

\bibitem[{\citenamefont{Vollhardt}(1993)}]{D.Vollhardt}
\bibinfo{author}{\bibfnamefont{D.}~\bibnamefont{Vollhardt}},
  \emph{\bibinfo{title}{{Lecture-Notes for the 9th Jerusalem Winter School for
  Theoretical Physics, Jerusalem 30. Dec. 1991 - 8. Jan. 1992, "Correlated
  Electron Systems"}}} (\bibinfo{publisher}{World Scientific},
  \bibinfo{address}{Singapore}, \bibinfo{year}{1993}), ISBN
  \bibinfo{isbn}{98102112321}.

\bibitem[{\citenamefont{Pruschke et~al.}(1995)\citenamefont{Pruschke, Jarrell,
  and Freericks}}]{Pruschke1995}
\bibinfo{author}{\bibfnamefont{T.}~\bibnamefont{Pruschke}},
  \bibinfo{author}{\bibfnamefont{M.}~\bibnamefont{Jarrell}}, \bibnamefont{and}
  \bibinfo{author}{\bibfnamefont{J.}~\bibnamefont{Freericks}},
  \bibinfo{journal}{Advances in Physics} \textbf{\bibinfo{volume}{44}},
  \bibinfo{pages}{187} (\bibinfo{year}{1995}), ISSN \bibinfo{issn}{0001-8732},
  \urlprefix\url{http://www.informaworld.com/openurl?genre=article&doi=10.1080%
/00018739500101526&magic=crossref||D404A21C5BB053405B1A640AFFD44AE3}.

\bibitem[{\citenamefont{Kotliar et~al.}(2006)\citenamefont{Kotliar, Savrasov,
  Haule, Oudovenko, Parcollet, and Marianetti}}]{Kotliar2006}
\bibinfo{author}{\bibfnamefont{G.}~\bibnamefont{Kotliar}},
  \bibinfo{author}{\bibfnamefont{S.}~\bibnamefont{Savrasov}},
  \bibinfo{author}{\bibfnamefont{K.}~\bibnamefont{Haule}},
  \bibinfo{author}{\bibfnamefont{V.}~\bibnamefont{Oudovenko}},
  \bibinfo{author}{\bibfnamefont{O.}~\bibnamefont{Parcollet}},
  \bibnamefont{and}
  \bibinfo{author}{\bibfnamefont{C.}~\bibnamefont{Marianetti}},
  \bibinfo{journal}{Reviews of Modern Physics} \textbf{\bibinfo{volume}{78}},
  \bibinfo{pages}{865} (\bibinfo{year}{2006}), ISSN \bibinfo{issn}{0034-6861},
  \urlprefix\url{http://link.aps.org/doi/10.1103/RevModPhys.78.865}.

\bibitem[{\citenamefont{Leonov et~al.}(2010)\citenamefont{Leonov, Korotin,
  Binggeli, Anisimov, and Vollhardt}}]{Leonov2009}
\bibinfo{author}{\bibfnamefont{I.}~\bibnamefont{Leonov}},
  \bibinfo{author}{\bibfnamefont{D.}~\bibnamefont{Korotin}},
  \bibinfo{author}{\bibfnamefont{N.}~\bibnamefont{Binggeli}},
  \bibinfo{author}{\bibfnamefont{V.~I.} \bibnamefont{Anisimov}},
  \bibnamefont{and}
  \bibinfo{author}{\bibfnamefont{D.}~\bibnamefont{Vollhardt}},
  \bibinfo{journal}{Physical Review B} \textbf{\bibinfo{volume}{81}},
  \bibinfo{pages}{075109} (\bibinfo{year}{2010}), ISSN
  \bibinfo{issn}{1098-0121}, \eprint{0909.1283v1},
  \urlprefix\url{http://link.aps.org/doi/10.1103/PhysRevB.81.075109}.

\bibitem[{\citenamefont{Kune\v{s} et~al.}(2010)\citenamefont{Kune\v{s},
  Baldassarre, Sch\"{a}chner, Rabia, Kuntscher, Korotin, Anisimov, McLeod,
  Kurmaev, and Moewes}}]{Wilson1971}
\bibinfo{author}{\bibfnamefont{J.}~\bibnamefont{Kune\v{s}}},
  \bibinfo{author}{\bibfnamefont{L.}~\bibnamefont{Baldassarre}},
  \bibinfo{author}{\bibfnamefont{B.}~\bibnamefont{Sch\"{a}chner}},
  \bibinfo{author}{\bibfnamefont{K.}~\bibnamefont{Rabia}},
  \bibinfo{author}{\bibfnamefont{C.~A.} \bibnamefont{Kuntscher}},
  \bibinfo{author}{\bibfnamefont{D.~M.} \bibnamefont{Korotin}},
  \bibinfo{author}{\bibfnamefont{V.~I.} \bibnamefont{Anisimov}},
  \bibinfo{author}{\bibfnamefont{J.~A.} \bibnamefont{McLeod}},
  \bibinfo{author}{\bibfnamefont{E.~Z.} \bibnamefont{Kurmaev}},
  \bibnamefont{and} \bibinfo{author}{\bibfnamefont{A.}~\bibnamefont{Moewes}},
  \bibinfo{journal}{Physical Review B} \textbf{\bibinfo{volume}{81}},
  \bibinfo{pages}{035122} (\bibinfo{year}{2010}), ISSN
  \bibinfo{issn}{1098-0121},
  \urlprefix\url{http://link.aps.org/doi/10.1103/PhysRevB.81.035122}.

\bibitem[{\citenamefont{Kune\v{s} et~al.}(2009)\citenamefont{Kune\v{s},
  Korotin, Korotin, Anisimov, and Werner}}]{Kunes2009}
\bibinfo{author}{\bibfnamefont{J.}~\bibnamefont{Kune\v{s}}},
  \bibinfo{author}{\bibfnamefont{D.~M.} \bibnamefont{Korotin}},
  \bibinfo{author}{\bibfnamefont{M.~A.} \bibnamefont{Korotin}},
  \bibinfo{author}{\bibfnamefont{V.~I.} \bibnamefont{Anisimov}},
  \bibnamefont{and} \bibinfo{author}{\bibfnamefont{P.}~\bibnamefont{Werner}},
  \bibinfo{journal}{Physical Review Letters} \textbf{\bibinfo{volume}{102}},
  \bibinfo{pages}{146402} (\bibinfo{year}{2009}), ISSN
  \bibinfo{issn}{0031-9007},
  \urlprefix\url{http://link.aps.org/doi/10.1103/PhysRevLett.102.146402}.

\bibitem[{\citenamefont{Leonov et~al.}(2011)\citenamefont{Leonov, Poteryaev,
  Anisimov, and Vollhardt}}]{Leonov2011}
\bibinfo{author}{\bibfnamefont{I.}~\bibnamefont{Leonov}},
  \bibinfo{author}{\bibfnamefont{A.~I.} \bibnamefont{Poteryaev}},
  \bibinfo{author}{\bibfnamefont{V.~I.} \bibnamefont{Anisimov}},
  \bibnamefont{and}
  \bibinfo{author}{\bibfnamefont{D.}~\bibnamefont{Vollhardt}},
  \bibinfo{journal}{Physical Review Letters} \textbf{\bibinfo{volume}{106}},
  \bibinfo{pages}{106405} (\bibinfo{year}{2011}), ISSN
  \bibinfo{issn}{0031-9007},
  \urlprefix\url{http://link.aps.org/doi/10.1103/PhysRevLett.106.106405}.

\bibitem[{\citenamefont{Cox and Sleight}(1976)}]{Cox1976969}
\bibinfo{author}{\bibfnamefont{D.~E.} \bibnamefont{Cox}} \bibnamefont{and}
  \bibinfo{author}{\bibfnamefont{A.~W.} \bibnamefont{Sleight}},
  \bibinfo{journal}{Solid State Communications} \textbf{\bibinfo{volume}{19}},
  \bibinfo{pages}{969} (\bibinfo{year}{1976}), ISSN \bibinfo{issn}{0038-1098},
  \urlprefix\url{http://www.sciencedirect.com/science/article/B6TVW-46Y385H-B/%
2/e12bd4afe5b397d5b92f00df04fb1803}.

\bibitem[{\citenamefont{Takagi et~al.}(1987)\citenamefont{Takagi, Uchida,
  Tajima, Kitazawa, and Tanaka}}]{TakagiHUchidaSTajimaS1987}
\bibinfo{author}{\bibfnamefont{H.}~\bibnamefont{Takagi}},
  \bibinfo{author}{\bibfnamefont{S.~I.} \bibnamefont{Uchida}},
  \bibinfo{author}{\bibfnamefont{S.}~\bibnamefont{Tajima}},
  \bibinfo{author}{\bibfnamefont{K.}~\bibnamefont{Kitazawa}}, \bibnamefont{and}
  \bibinfo{author}{\bibfnamefont{S.}~\bibnamefont{Tanaka}}, in
  \emph{\bibinfo{booktitle}{18th Int. Conf. on the Physics of Semiconductors}},
  edited by \bibinfo{editor}{\bibfnamefont{O.}~\bibnamefont{Engstr\"{o}m}}
  (\bibinfo{publisher}{Singapore: World Scientific},
  \bibinfo{address}{Stockholm}, \bibinfo{year}{1987}), pp.
  \bibinfo{pages}{1851--1855}.

\bibitem[{\citenamefont{Liechtenstein et~al.}(1991)\citenamefont{Liechtenstein,
  Mazin, Rodriguez, Jepsen, Andersen, and Methfessel}}]{PhysRevB.44.5388}
\bibinfo{author}{\bibfnamefont{A.~I.} \bibnamefont{Liechtenstein}},
  \bibinfo{author}{\bibfnamefont{I.~I.} \bibnamefont{Mazin}},
  \bibinfo{author}{\bibfnamefont{C.~O.} \bibnamefont{Rodriguez}},
  \bibinfo{author}{\bibfnamefont{O.}~\bibnamefont{Jepsen}},
  \bibinfo{author}{\bibfnamefont{O.~K.} \bibnamefont{Andersen}},
  \bibnamefont{and}
  \bibinfo{author}{\bibfnamefont{M.}~\bibnamefont{Methfessel}},
  \bibinfo{journal}{Physical Review B} \textbf{\bibinfo{volume}{44}},
  \bibinfo{pages}{5388} (\bibinfo{year}{1991}), ISSN \bibinfo{issn}{0163-1829},
  \urlprefix\url{http://link.aps.org/doi/10.1103/PhysRevB.44.5388}.

\bibitem[{\citenamefont{Franchini et~al.}(2009)\citenamefont{Franchini, Kresse,
  and Podloucky}}]{Franchini2009}
\bibinfo{author}{\bibfnamefont{C.}~\bibnamefont{Franchini}},
  \bibinfo{author}{\bibfnamefont{G.}~\bibnamefont{Kresse}}, \bibnamefont{and}
  \bibinfo{author}{\bibfnamefont{R.}~\bibnamefont{Podloucky}},
  \bibinfo{journal}{Physical Review Letters} \textbf{\bibinfo{volume}{102}},
  \bibinfo{pages}{256402} (\bibinfo{year}{2009}), ISSN
  \bibinfo{issn}{0031-9007},
  \urlprefix\url{http://link.aps.org/doi/10.1103/PhysRevLett.102.256402}.

\bibitem[{\citenamefont{Franchini et~al.}(2010)\citenamefont{Franchini, Sanna,
  Marsman, and Kresse}}]{Franchini2010}
\bibinfo{author}{\bibfnamefont{C.}~\bibnamefont{Franchini}},
  \bibinfo{author}{\bibfnamefont{A.}~\bibnamefont{Sanna}},
  \bibinfo{author}{\bibfnamefont{M.}~\bibnamefont{Marsman}}, \bibnamefont{and}
  \bibinfo{author}{\bibfnamefont{G.}~\bibnamefont{Kresse}},
  \bibinfo{journal}{Physical Review B} \textbf{\bibinfo{volume}{81}},
  \bibinfo{pages}{085213} (\bibinfo{year}{2010}), ISSN
  \bibinfo{issn}{1098-0121},
  \urlprefix\url{http://link.aps.org/doi/10.1103/PhysRevB.81.085213}.

\bibitem[{\citenamefont{Thonhauser and Rabe}(2006)}]{Thonhauser2006}
\bibinfo{author}{\bibfnamefont{T.}~\bibnamefont{Thonhauser}} \bibnamefont{and}
  \bibinfo{author}{\bibfnamefont{K.}~\bibnamefont{Rabe}},
  \bibinfo{journal}{Physical Review B} \textbf{\bibinfo{volume}{73}},
  \bibinfo{pages}{212106} (\bibinfo{year}{2006}), ISSN
  \bibinfo{issn}{1098-0121},
  \urlprefix\url{http://link.aps.org/doi/10.1103/PhysRevB.73.212106}.

\bibitem[{\citenamefont{Pei et~al.}(1990)\citenamefont{Pei, Jorgensen,
  Dabrowski, Hinks, Richards, Mitchell, Newsam, Sinha, Vaknin, and
  Jacobson}}]{PhysRevB.41.4126}
\bibinfo{author}{\bibfnamefont{S.}~\bibnamefont{Pei}},
  \bibinfo{author}{\bibfnamefont{J.~D.} \bibnamefont{Jorgensen}},
  \bibinfo{author}{\bibfnamefont{B.}~\bibnamefont{Dabrowski}},
  \bibinfo{author}{\bibfnamefont{D.~G.} \bibnamefont{Hinks}},
  \bibinfo{author}{\bibfnamefont{D.~R.} \bibnamefont{Richards}},
  \bibinfo{author}{\bibfnamefont{A.~W.} \bibnamefont{Mitchell}},
  \bibinfo{author}{\bibfnamefont{J.~M.} \bibnamefont{Newsam}},
  \bibinfo{author}{\bibfnamefont{S.~K.} \bibnamefont{Sinha}},
  \bibinfo{author}{\bibfnamefont{D.}~\bibnamefont{Vaknin}}, \bibnamefont{and}
  \bibinfo{author}{\bibfnamefont{A.~J.} \bibnamefont{Jacobson}},
  \bibinfo{journal}{Physical Review B} \textbf{\bibinfo{volume}{41}},
  \bibinfo{pages}{4126} (\bibinfo{year}{1990}), ISSN \bibinfo{issn}{0163-1829},
  \urlprefix\url{http://link.aps.org/doi/10.1103/PhysRevB.41.4126}.

\bibitem[{\citenamefont{Giannozzi et~al.}(2009)\citenamefont{Giannozzi, Baroni,
  Bonini, Calandra, Car, Cavazzoni, Ceresoli, Chiarotti, Cococcioni, Dabo
  et~al.}}]{Giannozzi2009}
\bibinfo{author}{\bibfnamefont{P.}~\bibnamefont{Giannozzi}},
  \bibinfo{author}{\bibfnamefont{S.}~\bibnamefont{Baroni}},
  \bibinfo{author}{\bibfnamefont{N.}~\bibnamefont{Bonini}},
  \bibinfo{author}{\bibfnamefont{M.}~\bibnamefont{Calandra}},
  \bibinfo{author}{\bibfnamefont{R.}~\bibnamefont{Car}},
  \bibinfo{author}{\bibfnamefont{C.}~\bibnamefont{Cavazzoni}},
  \bibinfo{author}{\bibfnamefont{D.}~\bibnamefont{Ceresoli}},
  \bibinfo{author}{\bibfnamefont{G.~L.} \bibnamefont{Chiarotti}},
  \bibinfo{author}{\bibfnamefont{M.}~\bibnamefont{Cococcioni}},
  \bibinfo{author}{\bibfnamefont{I.}~\bibnamefont{Dabo}}, \bibnamefont{et~al.},
  \bibinfo{journal}{Journal of Physics: Condensed Matter}
  \textbf{\bibinfo{volume}{21}}, \bibinfo{pages}{395502}
  (\bibinfo{year}{2009}), ISSN \bibinfo{issn}{0953-8984},
  \urlprefix\url{http://stacks.iop.org/0953-8984/21/i=39/a=395502?key=crossref%
.c21336c286fa6d3db893262ae3f6e151}.

\bibitem[{\citenamefont{Vanderbilt}(1990)}]{Vanderbilt1990}
\bibinfo{author}{\bibfnamefont{D.}~\bibnamefont{Vanderbilt}},
  \bibinfo{journal}{Physical Review B} \textbf{\bibinfo{volume}{41}},
  \bibinfo{pages}{7892} (\bibinfo{year}{1990}), ISSN \bibinfo{issn}{0163-1829},
  \urlprefix\url{http://link.aps.org/doi/10.1103/PhysRevB.41.7892}.

\bibitem[{\citenamefont{Monkhorst and Pack}(1976)}]{Monkhorst1976}
\bibinfo{author}{\bibfnamefont{H.}~\bibnamefont{Monkhorst}} \bibnamefont{and}
  \bibinfo{author}{\bibfnamefont{J.}~\bibnamefont{Pack}},
  \bibinfo{journal}{Physical Review B} \textbf{\bibinfo{volume}{13}},
  \bibinfo{pages}{5188} (\bibinfo{year}{1976}), ISSN \bibinfo{issn}{0556-2805},
  \urlprefix\url{http://link.aps.org/doi/10.1103/PhysRevB.13.5188}.

\bibitem[{\citenamefont{Anisimov and Gunnarsson}(1991)}]{Anisimov1991b}
\bibinfo{author}{\bibfnamefont{V.~I.} \bibnamefont{Anisimov}} \bibnamefont{and}
  \bibinfo{author}{\bibfnamefont{O.}~\bibnamefont{Gunnarsson}},
  \bibinfo{journal}{Physical Review B} \textbf{\bibinfo{volume}{43}},
  \bibinfo{pages}{7570} (\bibinfo{year}{1991}), ISSN \bibinfo{issn}{0163-1829},
  \urlprefix\url{http://link.aps.org/doi/10.1103/PhysRevB.43.7570}.

\bibitem[{\citenamefont{Anisimov et~al.}(2008)\citenamefont{Anisimov, Korotin,
  Streltsov, Kozhevnikov, Kune\v{s}, Shorikov, and Korotin}}]{Anisimov2008}
\bibinfo{author}{\bibfnamefont{V.~I.} \bibnamefont{Anisimov}},
  \bibinfo{author}{\bibfnamefont{D.~M.} \bibnamefont{Korotin}},
  \bibinfo{author}{\bibfnamefont{S.~V.} \bibnamefont{Streltsov}},
  \bibinfo{author}{\bibfnamefont{A.~V.} \bibnamefont{Kozhevnikov}},
  \bibinfo{author}{\bibfnamefont{J.}~\bibnamefont{Kune\v{s}}},
  \bibinfo{author}{\bibfnamefont{A.~O.} \bibnamefont{Shorikov}},
  \bibnamefont{and} \bibinfo{author}{\bibfnamefont{M.~A.}
  \bibnamefont{Korotin}}, \bibinfo{journal}{JETP Letters}
  \textbf{\bibinfo{volume}{88}}, \bibinfo{pages}{729} (\bibinfo{year}{2008}),
  ISSN \bibinfo{issn}{0021-3640}, \eprint{0807.0547v1},
  \urlprefix\url{http://www.springerlink.com/index/10.1134/S0021364008230069}.

\end{thebibliography}

\end{document}